\documentclass[aps,prd,twocolumn,superscriptaddress,showpacs,floatfix,nofootinbib]{revtex4}

\usepackage[dvips]{graphicx}
\usepackage{xspace}
\usepackage{latexsym}
\usepackage{amssymb}
\usepackage{times}
\usepackage{mathptm}

\newcommand{\pizero}[1]{\ensuremath{\pi^0}}              

\begin{document}

\title{Experimental study of the atmospheric neutrino backgrounds
for $p\rightarrow e^+\pi^0$ searches in water Cherenkov detectors}

\newcommand{\BCN}{\affiliation{Institut de Fisica d'Altes Energies, Universitat Autonoma de Barcelona, E-08193 Bellaterra (Barcelona), Spain}}
\newcommand{\BU}{\affiliation{Department of Physics, Boston University, Boston, Massachusetts 02215, USA}}
\newcommand{\UBC}{\affiliation{Department of Physics \& Astronomy, University of British Columbia, Vancouver, British Columbia V6T 1Z1, Canada}}
\newcommand{\UCI}{\affiliation{Department of Physics and Astronomy, University of California, Irvine, Irvine, California 92697-4575, USA}}
\newcommand{\SACLAY}{\affiliation{DAPNIA, CEA Saclay, 91191 Gif-sur-Yvette Cedex, France}}
\newcommand{\CNU}{\affiliation{Department of Physics, Chonnam National University, Kwangju 500-757, Korea}}
\newcommand{\DU}{\affiliation{Department of Physics, Dongshin University, Naju 520-714, Korea}}
\newcommand{\DUKE}{\affiliation{Department of Physics, Duke University, Durham, North Carolina 27708, USA}}
\newcommand{\GENEVA}{\affiliation{DPNC, Section de Physique, University of Geneva, CH1211, Geneva 4, Switzerland}}
\newcommand{\UH}{\affiliation{Department of Physics and Astronomy, University of Hawaii, Honolulu, Hawaii 96822, USA}}
\newcommand{\KEK}{\affiliation{High Energy Accelerator Research Organization(KEK), Tsukuba, Ibaraki 305-0801, Japan}}
\newcommand{\HIR}{\affiliation{Graduate School of Advanced Sciences of Matter, Hiroshima University, Higashi-Hiroshima, Hiroshima 739-8530, Japan}}
\newcommand{\INR}{\affiliation{Institute for Nuclear Research, Moscow 117312, Russia}}
\newcommand{\KOBE}{\affiliation{Kobe University, Kobe, Hyogo 657-8501, Japan}}
\newcommand{\KOR}{\affiliation{Department of Physics, Korea University, Seoul 136-701, Korea}}
\newcommand{\KYO}{\affiliation{Department of Physics, Kyoto University, Kyoto 606-8502, Japan}}
\newcommand{\LSU}{\affiliation{Department of Physics and Astronomy, Louisiana State University, Baton Rouge, Louisiana 70803-4001, USA}}
\newcommand{\MIT}{\affiliation{Department of Physics, Massachusetts Institute of Technology, Cambridge, Massachusetts 02139, USA}}
\newcommand{\MIYAGI}{\affiliation{Department of Physics, Miyagi University of Education, Sendai 980-0845, Japan}}
\newcommand{\NIIGATA}{\affiliation{Department of Physics, Niigata University, Niigata, Niigata 950-2181, Japan}}
\newcommand{\OKAYAMA}{\affiliation{Department of Physics, Okayama University, Okayama, Okayama 700-8530, Japan}}
\newcommand{\OSAKA}{\affiliation{Department of Physics, Osaka University, Toyonaka, Osaka 560-0043, Japan}}
\newcommand{\ROME}{\affiliation{University of Rome La Sapienza and INFN, I-000185 Rome, Italy}}
\newcommand{\SNU}{\affiliation{Department of Physics, Seoul National University, Seoul 151-747, Korea}}
\newcommand{\SOLTAN}{\affiliation{A.~Soltan Institute for Nuclear Studies, 00-681 Warsaw, Poland}}
\newcommand{\TOHOKU}{\affiliation{Research Center for Neutrino Science, Tohoku University, Sendai, Miyagi 980-8578, Japan}}
\newcommand{\SB}{\affiliation{Department of Physics and Astronomy, State University of New York, Stony Brook, New York 11794-3800, USA}}
\newcommand{\TUS}{\affiliation{Department of Physics, Tokyo University of Science, Noda, Chiba 278-0022, Japan}}
\newcommand{\KAM}{\affiliation{Kamioka Observatory, Institute for Cosmic Ray Research, University of Tokyo, Kamioka, Gifu 506-1205, Japan}}
\newcommand{\RCCN}{\affiliation{Research Center for Cosmic Neutrinos, Institute for Cosmic Ray Research, University of Tokyo, Kashiwa, Chiba 277-8582, Japan}}
\newcommand{\TRIUMF}{\affiliation{TRIUMF, Vancouver, British Columbia V6T 2A3, Canada}}
\newcommand{\VAL}{\affiliation{Instituto de F\'{i}sica Corpuscular, E-46071 Valencia, Spain}}
\newcommand{\UW}{\affiliation{Department of Physics, University of Washington, Seattle, Washington 98195-1560, USA}}
\newcommand{\WARSAW}{\affiliation{Institute of Experimental Physics, Warsaw University, 00-681 Warsaw, Poland}}

\BCN
\BU
\UBC
\UCI
\SACLAY
\CNU
\DU
\DUKE
\GENEVA
\UH
\KEK
\HIR
\INR
\KOBE
\KOR
\KYO
\LSU
\MIT
\MIYAGI
\NIIGATA
\OKAYAMA
\OSAKA
\ROME
\SNU
\SOLTAN
\TOHOKU
\SB
\TUS
\KAM
\RCCN
\TRIUMF
\VAL
\UW
\WARSAW

\author{S.~Mine}\UCI 
\author{J.~L.~Alcaraz}\BCN                
\author{S.~Andringa}\BCN 
\author{S.~Aoki}\KOBE 
\author{J.~Argyriades}\SACLAY 
\author{K.~Asakura}\KOBE 
\author{R.~Ashie}\KAM 
\author{F.~Berghaus}\UBC
\author{H.~Berns}\UW 
\author{H.~Bhang}\SNU 
% ...
\collaboration{The K2K Collaboration}\noaffiliation

\date{\today}

\begin{abstract}


%
%
The atmospheric neutrino background 
for proton decay via $p \rightarrow e^+\pi^0$ 
in ring imaging water Cherenkov detectors
is studied with an artificial accelerator neutrino beam
for the first time.
In total, 3.14$\times$10$^5$ neutrino events
corresponding to about 10~megaton-years
of atmospheric neutrino interactions
were collected by a 1,000 ton water Cherenkov detector~(KT).
%
The KT charged-current single $\pi^0$ production data 
are well reproduced by simulation programs
of neutrino and secondary hadronic interactions
used in the Super--Kamiokande~(SK) proton decay search.
The obtained $p\rightarrow e^+\pi^0$ background rate by the KT data
for SK from the atmospheric neutrinos whose energies are below 3~GeV is 
1.63 $^{+0.42}_{-0.33}$~(stat.) $^{+0.45}_{-0.51}$~(syst.)
(megaton-year)$^{-1}$.
This result is also relevant to possible future, 
megaton-scale water Cherenkov detectors.

\end{abstract}


\maketitle
\newpage



\section{Introduction}

Discovery of nucleon decay would constitute direct evidence for
grand unification of three fundamental forces
~\cite{gut1,gut2,gut3} and point the way to
a new theory beyond the standard model of elementary particle physics. 
Among many possible decay modes of nucleons,
the decay mode $p\rightarrow e^+\pi^0$ is dominant in a variety of such Grand
Unified Theories~(GUTs)~\cite{epi0th1,epi0th2,epi0th3}. 

The world's largest~(22.5~kiloton fiducial volume) water Cherenkov
detector experiment, Super--Kamiokande~(SK)~\cite{sk_detector}, has
set a stringent partial lifetime limit of $\tau/B_{p\rightarrow
  e^+\pi^0} > 5.4 \times 10^{33}$~years~(90\,\% C.L.) for proton decays
into $e^+$ and $\pi^0$~\cite{sk_epi0_exp1,sk_epi0_exp2,sk_epi0_exp3}
based on the observation of no candidates in an integrated exposure of
about 0.1~megaton-year~(Mtyr).  A $p\rightarrow e^+\pi^0$
signal would be clearly identified in SK as showering
Cherenkov rings corresponding to the positron and two gammas from
$\pi^0$ decay, 
with low net momentum and total invariant mass close to the proton mass.
The estimated detection efficiency for the signal is 40\,\%, with a
background rate of about 0.3~events
in the integrated exposure predicted by the Monte
Carlo simulation program~(MC). 
%

Characteristics of the background events
from atmospheric neutrinos to SK's $p\rightarrow e^+\pi^0$
search have been studied using MC~\cite{future_wc1}.
Charged-current~(CC) interactions of atmospheric $\nu_e$ with only
an electron and single $\pi^0$ visible in the final state 
are the dominant source of the background.
Neutral-current~(NC) interactions with only 
two $\pi^0$s visible in the final state are 
the remaining dominant background.
Parent neutrino energies between 1 and 3~GeV dominate
for both CC and NC background events.
In this study,
SK reports that about 24\,\% of the background events come from neutrinos
above 3~GeV.
The neutrino and final-state nuclear interactions are considered to be
the dominant uncertainties for the background rate estimation.
However, the background rate has been estimated only with MC 
without any quantitative systematic error estimation.

Some GUT models predict nucleon lifetimes just above the current limits, 
motivating proposals for future, megaton-scale SK-type water Cherenkov
detector experiments~\cite{koshiba_mton, future_wc1, future_wc2, future_wc3,
future_wc4, future_wc5, future_wc6}.
Since SK and these more massive experiments would observe a non-negligible 
number of background events in the future, it is essential to check the 
predicted background rates experimentally.

In this study,
neutrino and secondary hadronic interaction MCs
used to estimate the background rate in SK proton decay searches
are checked using muon neutrino beam data collected with a
1,000~ton water Cherenkov detector~(KT)
in the K2K long-baseline neutrino oscillation experiment.
The background rate for the $p\rightarrow e^+\pi^0$ mode
in SK and future megaton-scale water Cherenkov detectors is determined
using the KT data.

While the KT measures muon neutrino reactions rather than electron neutrino 
reactions,
the dynamics of pion production and re-scattering processes in
the oxygen nucleus are identical between the two neutrino flavors.
Therefore, rare CC $\nu_e$ interaction topologies 
which may mimic proton decay can be checked
using the KT $\nu_{\mu}$ events
with a muon and single $\pi^0$ visible in the final state.
The rate of the atmospheric CC $\nu_e$ background 
can be determined using KT CC single $\pi^0$ events
with corrections for differences in neutrino flux and detection efficiency
and the assumption that cross sections and final state kinematics 
for $\nu_e$ and $\nu_\mu$ are identical.
In addition, the rate of the NC background can be determined 
with two $\pi^0$s visible in the final state at the KT.

The number of $\nu_{\mu}$ interactions
in the KT is two orders of magnitude larger than previous
data sets used for similar background 
studies using other types of detectors
~\cite{prebtest1,prebtest2,prebtest3,prebtest4}.
The K2K beam is well
matched to the atmospheric neutrino spectrum and samples the energies around a few GeV where most backgrounds arise.
This is the first result of a proton decay background study
using accelerator neutrino beam data 
collected with a water Cherenkov detector.

The outline of this paper is as follows:
Section~II reviews the K2K neutrino beam and the KT detector.
Section~III-A describes the data sample in the KT and 
validates the neutrino interaction models.
Section~III-B calculates the $p\rightarrow e^+\pi^0$ background
rate for proton decay detectors.
Finally, Section~IV concludes
and summarizes these results.

\section{Experimental setup}

\subsection{K2K neutrino beam}

The wide-band neutrino beam used in the K2K long-baseline neutrino oscillation experiment~\cite{k2k_full}
was primarily muon flavored (about 97.3, 1.3, and 1.5\,\% for $\nu_{\mu}$, 
$\nu_e$, and $\bar{\nu}_\mu$, respectively)
with a mean energy of 1.3~GeV.
Protons accelerated by the KEK proton synchrotron to a
kinetic energy of 12~GeV were extracted in a single turn to the neutrino
beam line.  The duration of an extraction, or ``spill'', was 1.1~$\mu$sec.
The beam was 
transported to an aluminum target which was in the first of a pair of
horn magnets. The horn system focused positive pions produced in
proton-aluminum interactions into a 200~meter long decay volume.
A beam dump was located at
the end of the decay volume to absorb all surviving particles other than
neutrinos.
The neutrino beam's profile was measured using the distribution of
neutrino interaction vertices in a muon range detector~\cite{mrd},
one component of the K2K near neutrino detector system.
The energy and angular distributions of muons from CC
neutrino interactions were also continuously monitored.

\begin{figure}
  \includegraphics[width=8cm,clip]{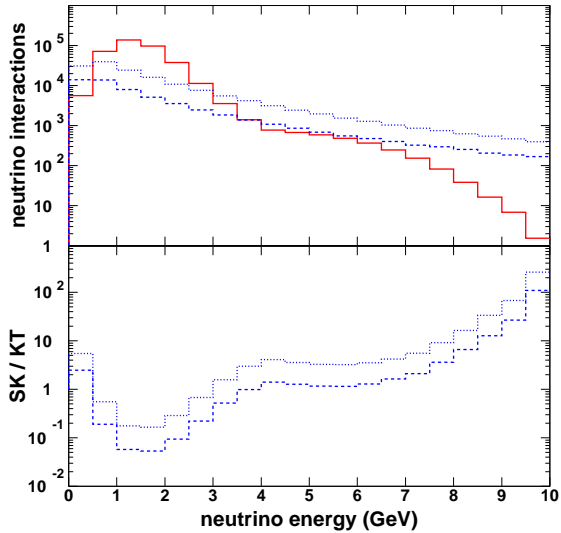}
  \caption
  {\protect \small The top figure compares the number of total $\nu_{\mu}$
    interactions in the KT 50~ton fiducial volume for
    7.4$\times$10$^{19}$ protons on target~(red solid) and
    the number of total neutrino interactions from atmospheric
    neutrinos in SK~\cite{honda, atmpd_full} for 1~Mtyr exposure~
    (blue dashed for $\nu_e$+$\bar{\nu}_e$ and blue dotted for
    $\nu_{all}$+$\bar{\nu}_{all}$, where $all$ stands for all the
    neutrino flavors).
    Disappearance of atmospheric $\nu_{\mu}$'s CC interactions due to
    neutrino oscillation is taken into account.
    The NEUT simulation program~\cite{neut} is used for calculation
    of total neutrino interaction for each neutrino flavor.
    The bottom figure shows the ratio of neutrino energy spectra for the
    atmospheric neutrino interactions in SK and the K2K beam $\nu_\mu$
    interactions in KT.  The meanings of the dashed and dotted lines are the
    same as above.}
  \label{flux_comp}
\end{figure}

Figure~\ref{flux_comp} compares total neutrino interaction 
(flux $\times$ total cross section $\times$ target volume $\times$ time)
spectra based on MC simulations 
for atmospheric neutrinos in SK~\cite{honda, atmpd_full}
with total $\nu_\mu$ interaction spectra for the K2K beam 
in the KT detector, located 300~m downstream of the proton target.
These total neutrino interaction 
spectra will be used for the background rate 
calculation shown in Sec.~III-B-1.
%
The neutrino spectrum measured by the KT covers the same energy range
as the portion of the atmospheric neutrino spectrum 
which dominates the production of proton decay background events.
Thus, analysis of the KT data allows a controlled study of the
neutrino interaction channels and the nuclear re-scattering processes
that determine the atmospheric neutrino backgrounds to nucleon decay
in water Cherenkov detectors.
In terms of the raw number of neutrino interactions between 1 and
3~GeV, the KT $\nu_{\mu}$ data correspond to about 10~Mtyr exposure to
the atmospheric $\nu_e$ flux at SK.  A more complete relationship
between the exposures will be derived in Sec.~III-B-2.

\subsection{The 1000~ton water Cherenkov detector}

The 1,000~ton ring imaging water Cherenkov detector~(KT) was located
in the K2K near detector hall.
The KT was designed as a smaller replica of SK,
using the same neutrino target material and instrumentation.
The inner volume of KT was a cylinder 8.6~m in diameter
and 8.6~m in height. This volume was viewed by 680 inward-facing
50~cm photomultiplier tubes (PMTs).  The PMTs and their arrangement
were identical to those of SK, covering 40\,\% of the detector's inner 
surface with active photo-cathode.
The software-defined fiducial volume for this analysis, based on the
reconstructed vertex for each neutrino interaction, is 50~tons in a
4~m by 4~m cylinder at the center of the tank, oriented along the beam
axis.  This fiducial volume is twice the standard fiducial volume
(25~tons) used in other KT analyses~\cite{k2k_full,1kt_pi0} in order
to increase the statistics of the data set.  Event selection within the
fiducial volume results in a pure neutrino sample, with negligible
contamination from cosmic rays and beam-induced muons entering the KT.

The KT data acquisition system was also similar to that of SK.  The
charge and timing information for each PMT hit above a threshold of
approximately 1/4 photo-electron was digitized by custom electronics
modules developed for SK~\cite{atm}.
The KT detector was triggered if there were more than 40 hit PMTs
in a 200~ns timing window during the spill gate. 
The trigger threshold was roughly equivalent to the
signal of a 6~MeV electron.  The analog sum of all 680 PMTs' signals~(PMTSUM)
was also recorded during every spill by a 500~MHz FADC,
to identify multiple neutrino interactions in a single spill.
The number of interactions in a spill was determined by counting
peaks in PMTSUM above a threshold equivalent to a 100~MeV electron.
Only events with a single neutrino interaction in the spill, as
determined by the PMTSUM peak search, are used in this study.

The KT and SK share a common library of event reconstruction
algorithms~\cite{shio_nim} and detector simulation programs based on
GEANT3~\cite{geant}.  
Similar detector calibration procedures are used in the KT~\cite{k2k_full} 
and SK, and similar detector performances are observed.

\subsubsection{Event reconstruction performance} 

The KT event reconstruction performance parameters which are most
relevant to this study are presented in this sub-section.
These reconstructions are used to select events and
results of these performance studies are used to estimate systematic errors 
in this study.

A relativistic charged particle emits Cherenkov light in a cone with
opening angle $\cos^{-1}(1/n\beta) \sim 41^{\circ}$, where $n \sim
1.34$ is the refractive index for water, aligned with the particle's
direction, and imaged as a ring by the detector's PMT grid.  
Electrons and gammas generate electromagnetic showers
(electrons also suffer multiple scatterings),
which produce diffused ring patterns, 
while muons and charged pions create sharper ring images, 
with well-defined edges.

The interaction vertex for each event is
reconstructed using PMT timing information.  After fixing the vertex,
the number of Cherenkov rings and their directions are determined by a
maximum-likelihood procedure.  The particle identification~(PID)
algorithm uses the charge pattern to classify each ring as a
showering, electron-like particle~($e^{\pm}$, $\gamma$) or a
non-showering, muon-like particle~($\mu^{\pm}$, $\pi^{\pm}$). For
events with a single ring, charge information is used to refine the
reconstructed vertex based on the PID result.
%
%
%
Finally, the particle momentum of each ring is determined using charge information,
based on the PID classification of the ring.

\begin{figure}
  \includegraphics[width=8cm,clip]{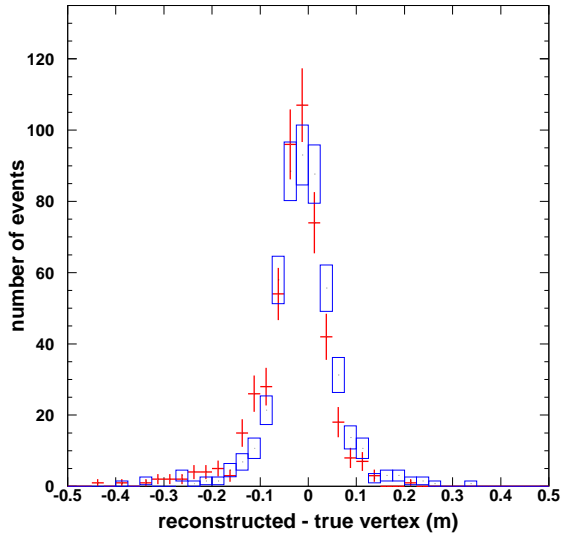}
  \caption
  {\protect \small Reconstructed - true vertex position of ``cosmic
    ray pipe'' muons entering at the center of KT.  Red crosses~(blue
    boxes) show the data~(MC).}
  \label{kt_crp}
\end{figure}

Vertex reconstruction in the KT was checked {\it in situ} with a special device
called a ``cosmic ray pipe'', which is a 25~cm diameter PVC pipe with
scintillation counters at both ends.  It was inserted from the top of
the tank through a calibration access port near the center of the
detector's top wall.  A coincidence between the two scintillation
counters tagged a cosmic ray muon entering the detector at the lower
end of the pipe.  Such a muon has a known direction along the pipe and
will begin radiating Cherenkov light when it leaves the pipe at a
known location inside the detector.  The ``cosmic ray pipe'' muon data
were taken at five vertical positions: 0, $\pm$1, and $\pm$2~m from
the center of the fiducial volume.  Figure~\ref{kt_crp} shows a typical
reconstructed vertex distribution of ``cosmic ray pipe'' muons.
The vertex resolution along the vertical direction was 10~cm or
better, and the central values of the distributions for data and MC
agree within 4~cm for all test points.  
The events in Fig.~\ref{kt_crp} have had a vertex refinement
based on charge and PID information applied to them.
%
The vertex resolution for multi-ring decays without this refinement,
such as simulated $p\rightarrow\mu^+\pi^0$ signal events, is 28~cm.
%

\begin{figure}
  \includegraphics[width=8cm,clip]{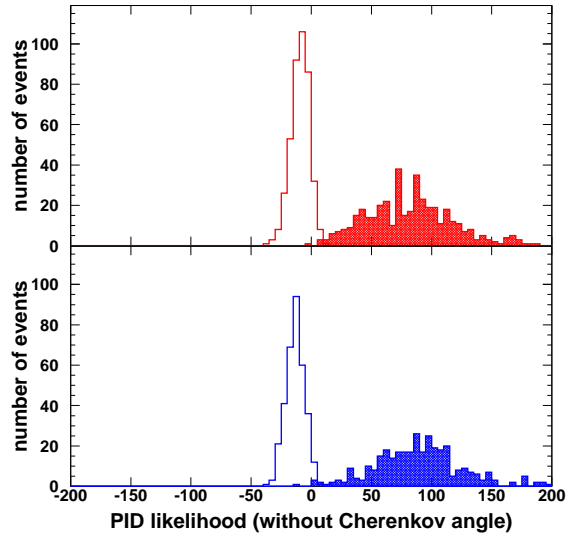}
  \caption
  {\protect \small PID likelihood 
    for cosmic ray muons~(shaded histogram) and the decay
    electrons~(open histogram).  
    The red~(blue) histogram~(top~(bottom) figure) shows data~(MC).  
    Positive~(negative) likelihood corresponds to
    muon~(electron)-like.}
  \label{kt_pid}
\end{figure}

The PID performance is checked with cosmic ray muons
and their associated decay electrons. 
As shown in Fig.~\ref{kt_pid},
the PID likelihood distributions are clearly separated
between the two types of rings,
although note that
the momenta of particles from the proton decay is typically
higher than that of Michel electrons.  Mis-ID probabilities for
500~MeV/$c$ muons and electrons were estimated with Monte Carlo to be
0.5\,\% and 0.8\,\%, respectively.

\begin{figure}
  \includegraphics[width=8cm,clip]{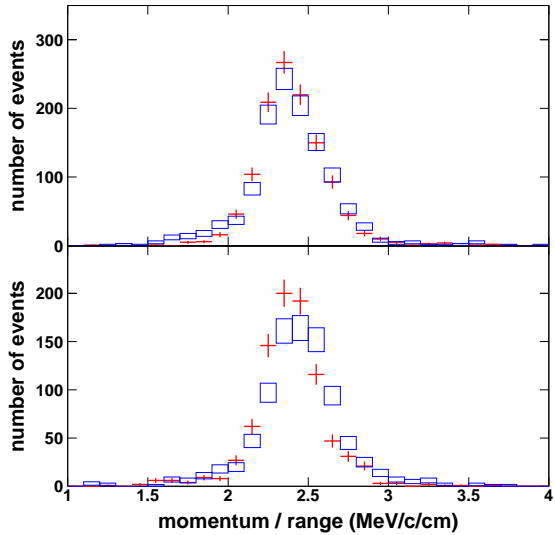}
  \caption
  {\protect \small Reconstructed momentum divided by range for
    vertical~(top figure) and horizontal~(bottom) cosmic ray muons.
    Red crosses~(blue boxes) show the data~(MC).}
  \label{kt_stmu_mom}
\end{figure}

\begin{figure}
  \includegraphics[width=8cm,clip]{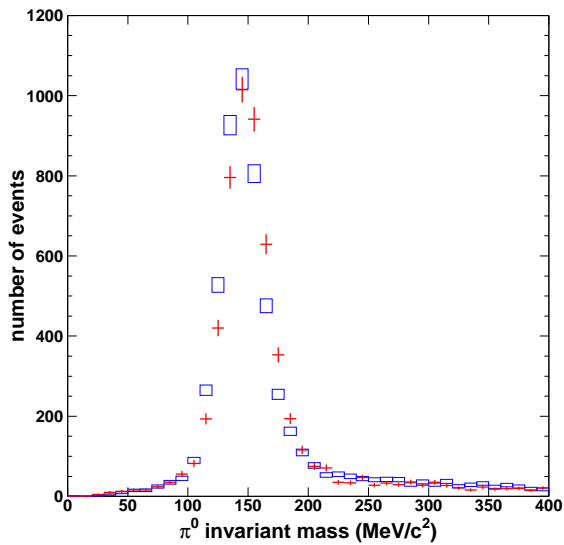}
  \caption
  {\protect \small Reconstructed mass of $\pi^0$ candidates produced
    by NC interactions in the K2K beam. Red crosses~(blue boxes) show
    the data~(MC).}
  \label{kt_pi0_mass}
\end{figure}

Figure~\ref{kt_stmu_mom} shows momentum loss,
as estimated by reconstructed momentum divided by measured range, 
for stopping cosmic ray muons.
The stopping point of a muon is determined by the reconstructed vertex
of its decay electron.  
Both vertical stopping muons which enter from the top
of the tank toward the bottom direction and horizontal stopping muons
entering from one side of the barrel toward the other are shown in
Fig.~\ref{kt_stmu_mom}.  The largest difference~(3\,\%) between data
and MC comes from the horizontal muons, and is used in the estimate of
the absolute energy scale uncertainty. The difference~(1.7\,\%) between
vertical and horizontal muons is used to estimate the detector
asymmetry of the energy scale.  The directional dependence of
reconstructed momentum of Michel electrons is also checked and found
to be uniform within the statistical error~(2.4\,\%) of the sample.
Figure~\ref{kt_pi0_mass} shows the reconstructed invariant mass of
NC single-$\pi^0$ events induced by the neutrino beam~\cite{1kt_pi0}.
The events must be fully-contained (deposit all of their Cherenkov light
inside the inner detector) and have two electron-like rings.
The $\pi^0$ mass peak is clearly observed, but shifted slightly higher
than the nominal value of 135~MeV/c$^2$.  This shift is the result of
energy deposit from de-excitation $\gamma$'s from oxygen nuclei in
neutrino interactions as well as a bias in the $\gamma\gamma$ opening
angle due to known vertex reconstruction bias. Each effect is several
MeV and is included in MC.
The remaining difference between data and MC is 4\,\% by comparing the
$\pi^0$ mass peaks, and is also used to estimate the absolute energy
scale uncertainty.  Taking these measurements together, the absolute
energy scale uncertainty is determined to be $^{+3}_{-4}$\,\% from the
horizontal stopping cosmic ray muon and $\pi^0$ samples, and the
detector asymmetry of the energy scale is 1.7\,\%.  The absolute energy
scale was stable within 1\,\% over the period when data for the present
study were collected.

\section{Data analysis}

All available good quality KT data from the 2000--2004 K2K running periods 
were used in this analysis.  In this period, 3.14$\times$10$^5$ total neutrino
events (2.75$\times$10$^5$ events with a single neutrino interaction
in a spill) in the 50 ton fiducial volume, corresponding to 
7.4 $\times$ 10$^{19}$~protons on target~(pot), were collected.

\subsection{Validation of neutrino interaction simulations}

Several packages to simulate neutrino interactions have been
developed~\cite{numc}. Both the NEUT~\cite{neut} and
NUANCE~\cite{nuance} simulation programs have been used in SK to
estimate the background rates for nucleon decay searches.  These
programs have already been used for various physics analyses and
confirmed to reproduce data well~\cite{k2k_full, atmpd_full, 1kt_pi0}.
In this analysis, the neutrino-induced $\mu^+\pi^0$
sample is used to do a careful comparison of the rates and distributions
relevant to the $p\rightarrow e^+\pi^0$ search.

\subsubsection{Data and MC event selection criteria}

CC $\nu_{\mu}$ interactions producing only a 
muon and single $\pi^0$ visible in the
final state~(``$\mu\pi^0$'' events) can,
except for the choice of lepton PID, be selected in the KT
with nearly the same cuts 
used for the SK $p\rightarrow e^+\pi^0$ search:
\newcounter{Lcount}
\begin{list}{(\Alph{Lcount})}
{\usecounter{Lcount} 
\setlength{\rightmargin}{\leftmargin}}
\item {\it Fully-contained~(FC):}  \\
The FC criterion in the KT detector requires that
no PMT signal greater than 200~photo-electrons is recorded,
since charged particles leaving the detector produce a large signal
in the PMT around their exit point.
\item {\it Two or three identified rings:} \\
At the neutrino energies relevant to this study, recoiling protons
are usually invisible because they are below Cherenkov threshold.  
Events with both two and three rings are accepted because, in some
cases, one of the two gammas from the $\pi^0$ decay is missed during
reconstruction.  This can happen when the decay is very asymmetric
in energy, or the rings are too close to resolve.
\item {\it PID:} \\
One identified ring must be muon-like and 
the other one or two rings must be electron-like.
\item {\it $\pi^0$ mass:} \\
For three-ring events, the reconstructed invariant mass of
the two electron-like rings must be between 85 and 215~MeV/$c^2$.
\end{list}

The SK $p \rightarrow e^+ \pi^0$ selection criteria are identical,
except that
the outer detector information is used to select the FC events,
all rings must be electron-like, 
and events with one or more identified muon decays are rejected.

Finally, the proton decay signal box defines a ``$p \rightarrow \mu\pi^0$'' sub-sample
of ``$\mu \pi^0$'' events consistent with the proton mass (about 1~GeV/$c^2$)
and Fermi motion in oxygen.
The oxygen nucleus is modeled with as a relativistic Fermi gas
with the Fermi surface momentum set to about 220~MeV/c.
The selection criteria used are:
\begin{list}{(\Alph{Lcount})}
{\usecounter{Lcount}
\setcounter{Lcount}{4}
\setlength{\rightmargin}{\leftmargin}}
\item {\it Proton decay kinematics:} \\
Total momentum $P_{tot}$ must be less than 250~MeV/$c$ and 
total invariant mass $M_{tot}$ must be between 800 and 1050~MeV/$c^2$, where:
\end{list}
\begin{itemize}
\item \(P_{tot} \equiv |\sum_i \vec p_i|\), 
\item \(M_{tot} \equiv \sqrt{E_{tot}^2 - P_{tot}^2} \),
\item \(E_{tot} \equiv \sum_{i} \sqrt{|\vec p_i|^2 + m_i^2}\),
\item \(\vec p_i\) is the reconstructed momentum vector of the \(i\)-th ring, and 
\item \(m_i\) is the mass of 
the particle corresponding to the \(i\)-th ring inferred from PID (either gamma or muon).
\end{itemize}

\begin{figure}
          \includegraphics[width=8cm,clip]{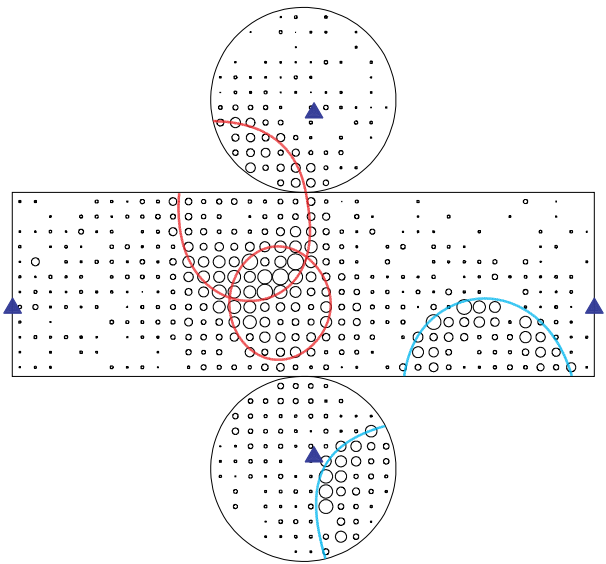}
          \caption
         {\protect \small Event display of ``$p \rightarrow \mu\pi^0$''
sample in the KT real data satisfying selection criteria (A)--(E).  
Each hit PMT is represented by
          a small circle whose area is proportional to the PMTs measured charge.
          The small blue triangles show the projection of the reconstructed vertex onto
          the walls.  Cyan~(red) thick circles show reconstructed rings identified
          by the PID as muon~(electron)-like (Two overlapped thick circles
          are identified as electron-like rings).
          The reconstructed $M_{\pi^0}$, $M_{tot}$, and $P_{tot}$
          for this event are $164~\hbox{MeV}/c^2$, $851~\hbox{MeV}/c^2$, and $226~\hbox{MeV}/c$,
          respectively.}
          \label{kt_event_display}
          \end{figure}

Figure~\ref{kt_event_display} shows the event display
of ``$p \rightarrow \mu\pi^0$'' sample in the KT real data 
falling within the proton decay signal box.
Three Cherenkov rings are clearly visible and well reconstructed.

\subsubsection{MC data samples}

MC $\nu_\mu$ data samples corresponding to $1.1 \times 10^{20}$~pot
and $5.4 \times 10^{19}$~pot were generated by NEUT and NUANCE,
respectively, and fully reconstructed after full simulation of
detector response using GEANT.  These MC samples are used to
generally study the ``$\mu\pi^0$'' events.  To extensively investigate
events in the proton decay signal box, additional $\nu_{\mu}$
interactions were generated with NEUT for 1.4, 5.0, 32.4, 54.0, and
$35.9 \times 10^{21}$~pot in neutrino energy bins 0--2.5, 2.5--3,
3--4, 4--5, and 5--6~GeV, respectively.  This additional sample was
preselected before reconstruction, using loose cuts on the total
charge and charge anisotropy from the generated vertex.
The breakdown of events surviving the selection criteria into
different reaction channels for the NEUT sample is summarized in
Tbl.~\ref{neutmode}.  Resonant CC single pion production dominates
both the ``$\mu \pi^0$'' and ``$p \rightarrow \mu \pi^0$'' samples.  The
quasi-elastic contribution arises through production of $\pi^0$ by
recoil nucleon interactions with H$_2$O (also simulated by the GEANT
and CALOR~\cite{calor} packages).
Identification of charged pions as muons by the PID is the dominant source
of the NC fraction.

  \begin{table}
  \begin{center}
  \begin{tabular}{lcc}
  \hline
  Event category & (A)-(D)~(\%) & (A)-(E)~(\%)\\
  & ``$\mu\pi^0$'' sample & ``$p \rightarrow \mu \pi^0$'' sample\\
  \hline
  CC quasi elastic & 12.2$\pm$0.2~(4.3$\pm$0.2) & 18.6$\pm$1.9 \\
  CC single pion from resonance & 50.2$\pm$0.3~(48.7$\pm$0.6) & 60.4$\pm$2.4 \\
  CC multi pions & 22.2$\pm$0.3~(29.9$\pm$0.5) & 16.1$\pm$1.8\\
  CC deep inelastic scattering & 0.6$\pm$0.1~(0.5$\pm$0.1) & 1.0$\pm$0.5 \\
  NC & 14.8$\pm$0.2~(16.5$\pm$0.4) & 4.0$\pm$1.0\\
  \hline
  \end{tabular}
  \end{center}
  \caption{\protect \small
   NEUT abundance of different reaction channels~\cite{neut}
   after applying event selections (A)--(D) and (A)--(E).
   Note that hadron system invariant mass $W$ is different for "multi pions" 
   and "deep inelastic scattering" modes as 1.3$\le W \le$2.0~GeV and 
   $W \ge$2.0~GeV, respectively. 
  "NC" includes all the modes shown for "CC". 
   The errors shown are MC statistical.  The number in parentheses for (A)--(D)
   shows the population for events with three identified rings only.}
   \label{neutmode}
   \end{table}

\begin{figure}
          \includegraphics[width=8cm,clip]{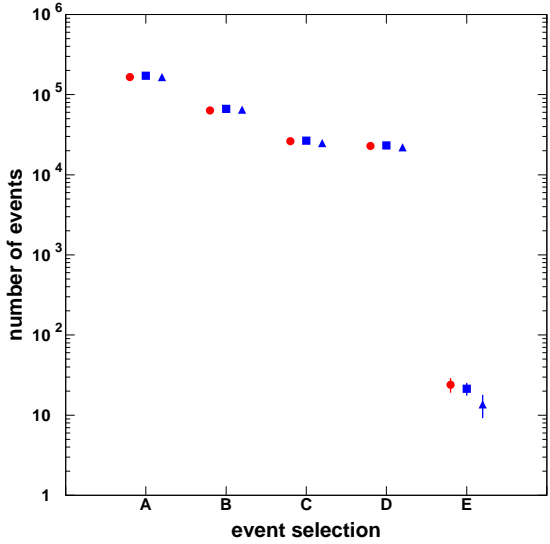}
          \caption
         {\protect \small Event rate after application of each selection criterion.
          The event selections (A)-(E) are explained in text.
	  The event rates at D and E correspond to
          ``$\mu\pi^0$'' and ``$p \rightarrow \mu \pi^0$'' samples,
          respectively.
          The data, NEUT, and NUANCE are shown as red circles,
          blue squares, and blue triangles, respectively.
          The simulated data are normalized by the total number of neutrino 
          interactions in the 25~ton fiducial volume.}
          \label{kt_event_rate}
          \end{figure}

Figure~\ref{kt_event_rate} shows that the rates of events surviving
after each selection (A--E) for data and both MC samples agree well.
The simulated data are normalized by the number of total neutrino
interactions in the 25~ton fiducial volume~\cite{k2k_full} in this plot.

For the data, 24 events remain in the proton decay signal box
for 7.4 $\times$ 10$^{19}$~pot,
while $21.4 \pm 3.8~\hbox{(stat)}$ and $13.6 \pm 4.4~\hbox{(stat)}$
are expected from NEUT and NUANCE, respectively, 
for the same number of total neutrino interactions
in the 25~ton fiducial volume.

All simulated events in the proton decay signal box were inspected,
and about 90\,\% show a correctly reconstructed, back-to-back muon and $\pi^0$
from the CC neutrino interactions.
The remaining 10\,\% mostly result from identification of a proton
as a gamma, or a charged pion as a gamma or a muon.
This remaining fraction is used to estimate the systematic error
on the efficiency difference between SK and KT shown in Tbl.~II.

In the MC,
98\,\% of ``$p \rightarrow \mu \pi^0$'' events found in the signal box are
from neutrinos interacting with nucleons in the oxygen nucleus. 
The remaining 2\,\% of events are from neutrinos interacting with free protons
where a $\pi^0$ is produced by hadronic interaction of 
the struck nucleon with H$_2$O.

\subsubsection{Comparison of kinematics between data and MC}

Since the background for proton decay searches depends not only on the
neutrino interaction rate but also on the event topologies, it is
important to verify that the simulations correctly reproduce the
kinematics of each particle in the final state.
Resonant single-pion production, which is the dominant source of the
background, can be characterized by the momentum transfer $Q^2$ and
the invariant mass of the hadronic system $W$.  Assuming a
resonance-mediated three-body reaction ($\nu_\mu n \rightarrow \mu^-
A, A \rightarrow p \pi^0$) and neglecting the Fermi motion and binding
energy of the target, the kinematic variables of interest
can be reconstructed from the observable quantities:
\begin{widetext}
\begin{displaymath}
W^2 = \frac{ 
( E_{\pi} - p_{\pi} \cos\theta_{\pi} ) 
( 2 m_n E_{\mu} - m_n^2 - m_{\mu}^2 ) +
  ( 2 m_n E_{\pi} - 2 E_{\mu} E_{\pi} + 
2 p_{\mu} p_{\pi} \cos\theta_{\mu\pi} + m_p^2 - m_{\pi}^2 )
( m_n - E_{\mu} + p_{\mu} \cos\theta_{\mu} ) 
} 
{
( m_n - E_{\mu} + p_{\mu} \cos\theta_{\mu} 
- E_{\pi} + p_{\pi} \cos\theta_{\pi} )},
\end{displaymath}
\end{widetext}
and
\begin{displaymath}
Q^2 = 2 E_{\nu} ( E_{\mu} - p_{\mu} \cos\theta_{\mu} ) - m_{\mu}^2,
\end{displaymath}
in natural units where $p$, $\theta$, $m$, and $E$ are a given
particle's reconstructed momentum, reconstructed angle from the beam
direction, mass, and calculated energy, respectively,
and $\theta_{\mu\pi}$ as the angle between reconstructed muon and $\pi^0$
directions. Also,
\begin{displaymath}
E_{\nu} = \frac{2 m_n E_{\mu} + W^2 - m_n^2 - m_{\mu}^2 }
{2( m_n - E_{\mu} + p_{\mu} \cos\theta_{\mu} )}
\end{displaymath}
is the inferred neutrino energy where ``$\pi$'', ``$n$'', ``$\mu$'', and
``$p$'' in these formulas stand for $\pi^0$, neutron, $\mu^-$, and
proton, respectively.

\begin{figure}
          \includegraphics[width=8cm,clip]{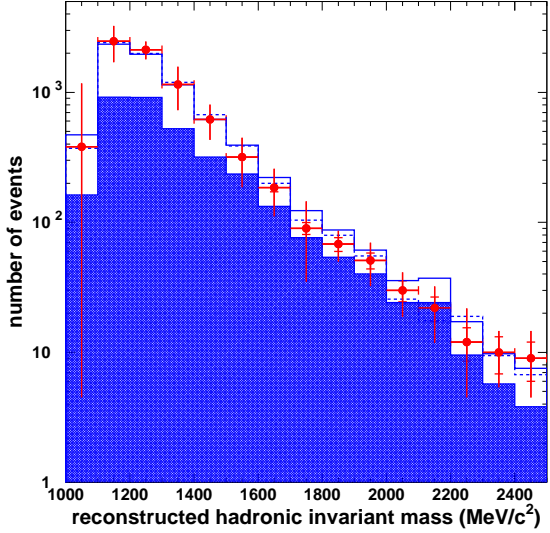}
          \caption
         {\protect \small
           Reconstructed hadronic invariant mass $W$
           for three ring ``$\mu\pi^0$'' events,
           assuming CC single $\pi^0$ production via resonance.
           Red crosses show the data with statistical and total measurement errors.
 	    The relatively large systematic errors in this plot
           are correlated between bins, and
           arise from the absolute energy scale uncertainty.
           The NEUT~(NUANCE) prediction is shown by the solid~(dashed) blue histogram.
           The hatched histogram shows the distribution of NEUT events that
           originate from resonant CC single pion production.
           Both MC samples are normalized to the data by number of entries.}
          \label{kt_mhad}
          \end{figure}

\begin{figure}
          \includegraphics[width=8cm,clip]{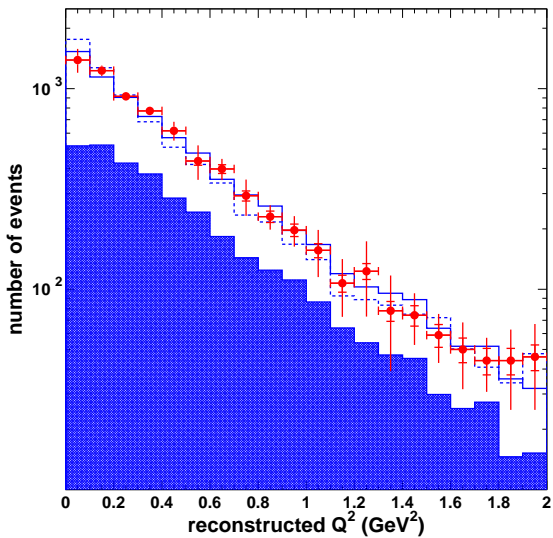}
          \caption
         {\protect \small
           Reconstructed momentum transfer $Q^2$ for three ring ``$\mu\pi^0$'' events,
           assuming CC single $\pi^0$ production via resonance.
           Red crosses show the data with statistical and total measurement errors.
           The NEUT~(NUANCE) prediction is shown by the solid~(dashed) blue histogram.
           The hatched histogram shows the distribution of NEUT events that
           originate from resonant CC single pion production.
           Both MC samples are normalized to the data by number of entries.}
          \label{kt_q2}
          \end{figure}

Figs.~\ref{kt_mhad} and~\ref{kt_q2} show
the reconstructed hadronic mass $W$ and momentum transfer $Q^2$ 
of three ring ``$\mu\pi^0$'' events.
Uncertainties in the KT event reconstruction are
accounted for in the real data points as correlated systematic errors,
so the error bars shown on each bin are not independent.
Although there may be a slight excess at small $Q^2$ in NUANCE,
the agreement between data and both MC samples is good 
for both kinematic variables within  the measurement errors.



The kinematic variables $Q^2$ and $W$ do not translate directly into
the quantities $M_{tot}$ vs. $P_{tot}$ used as the final selection for
proton decay candidates, so it is important to check the latter as well.
Figure~10-12 show
the distributions of $P_{tot}$ vs. $M_{tot}$ for all ``$\mu\pi^0$'' events 
from data,
NEUT, and NUANCE, respectively.

\begin{figure}
  \includegraphics[width=8cm,clip]{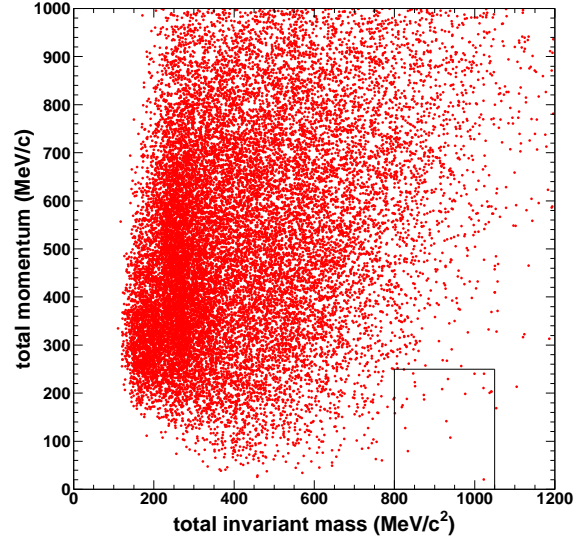}
  \caption
  {\protect \small Total momentum $P_{tot}$ vs. total invariant mass
    $M_{tot}$ for ``$\mu\pi^0$'' events from the KT data~(7.4 $\times$
    10$^{19}$~pot), with the proton decay signal box superimposed.}
  \label{kt_scat_data}
\end{figure}
\begin{figure}
  \includegraphics[width=8cm,clip]{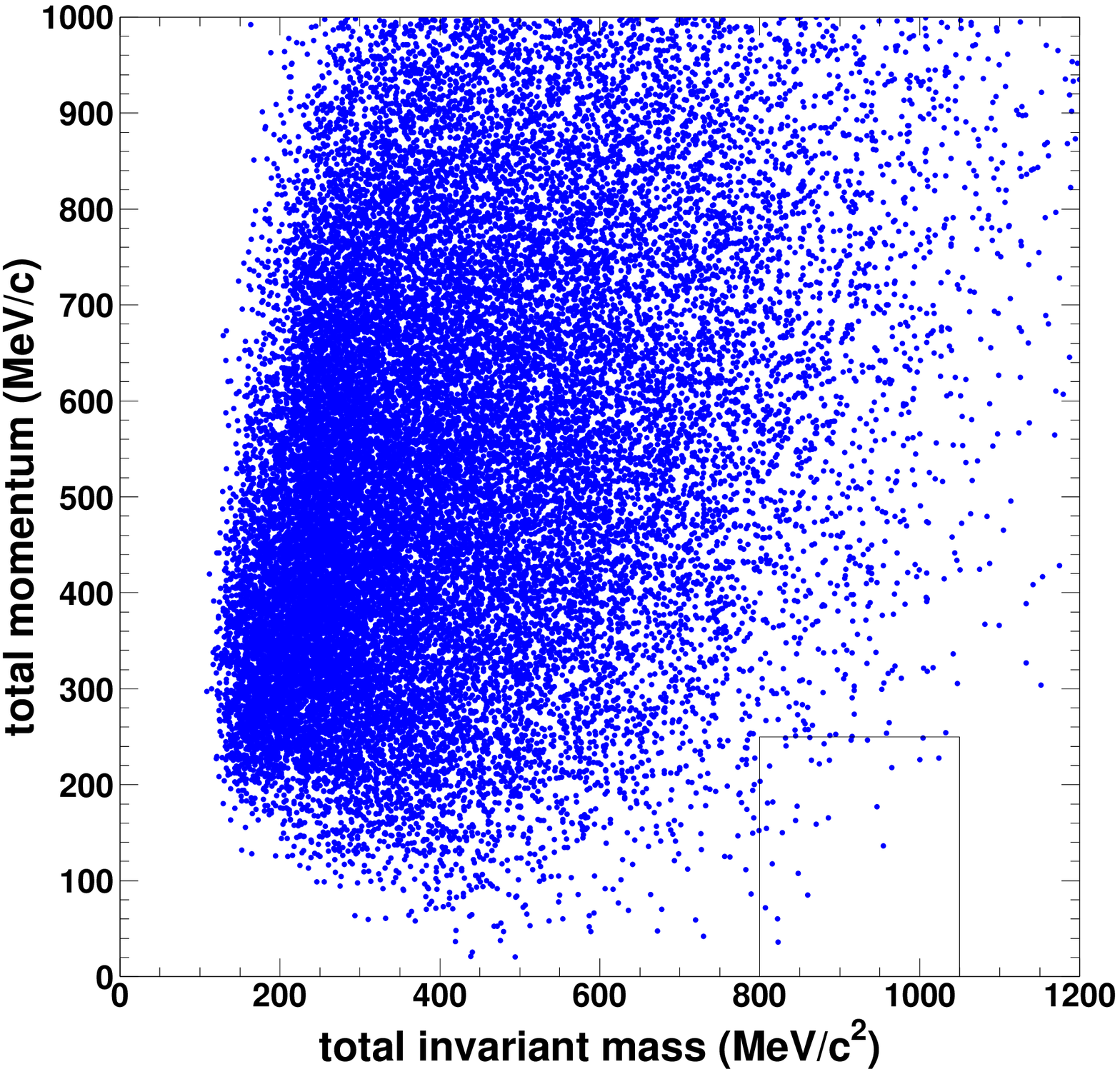}
  \caption
  {\protect \small Total momentum $P_{tot}$ vs. total invariant mass
    $M_{tot}$ for ``$\mu\pi^0$'' events from NEUT~($1.1 \times
    10^{20}$~pot), with the proton decay signal box superimposed.}
  \label{kt_scat_mc}
\end{figure}
\begin{figure}
  \includegraphics[width=8cm,clip]{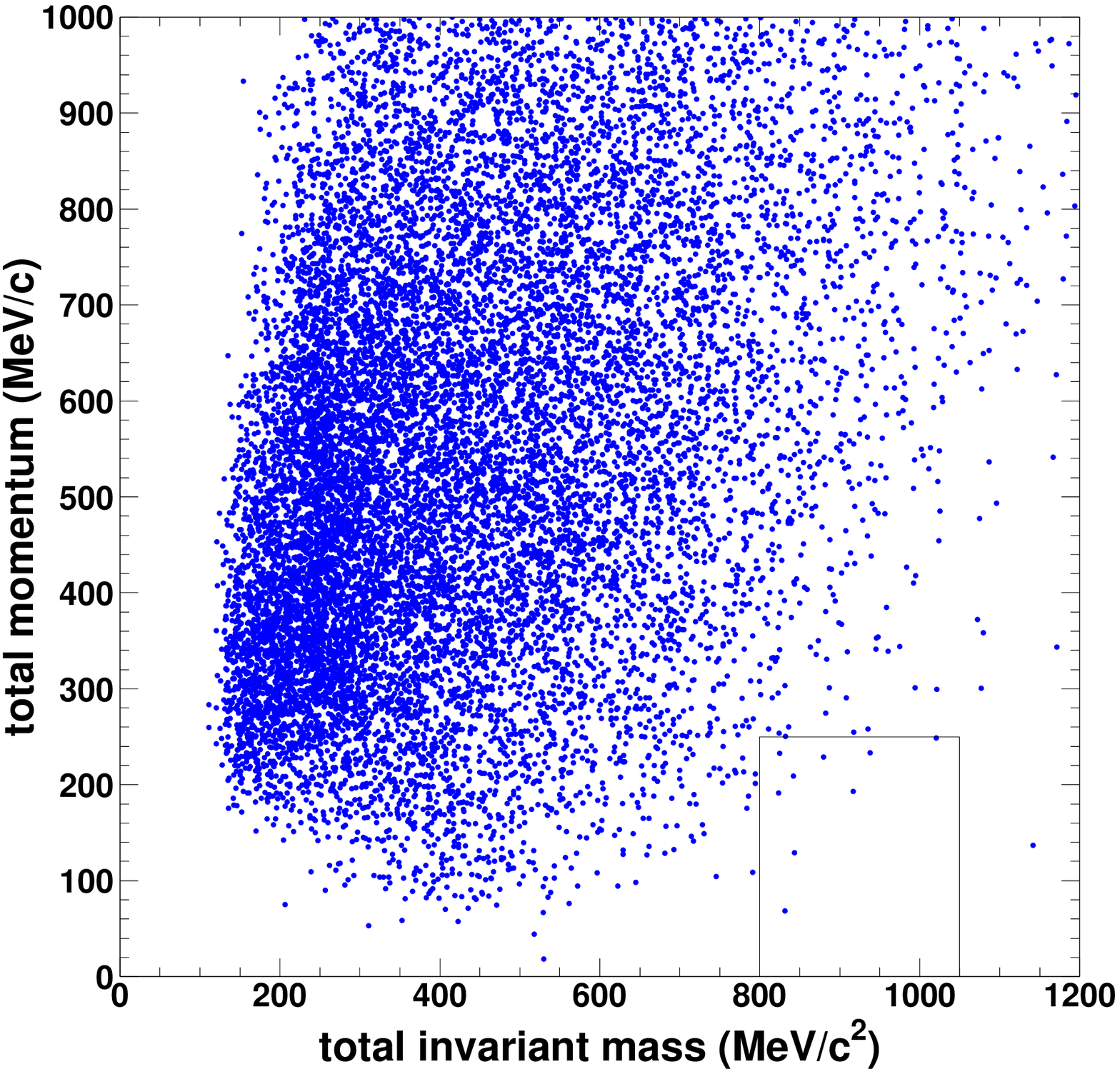}
  \caption
  {\protect \small Total momentum $P_{tot}$ vs. total invariant mass
    $M_{tot}$ for ``$\mu\pi^0$'' events from NUANCE~($5.4 \times
    10^{19}$~pot), with the proton decay signal box superimposed.}
  \label{kt_scat_nua}
\end{figure}

\begin{figure}
          \includegraphics[width=8cm,clip]{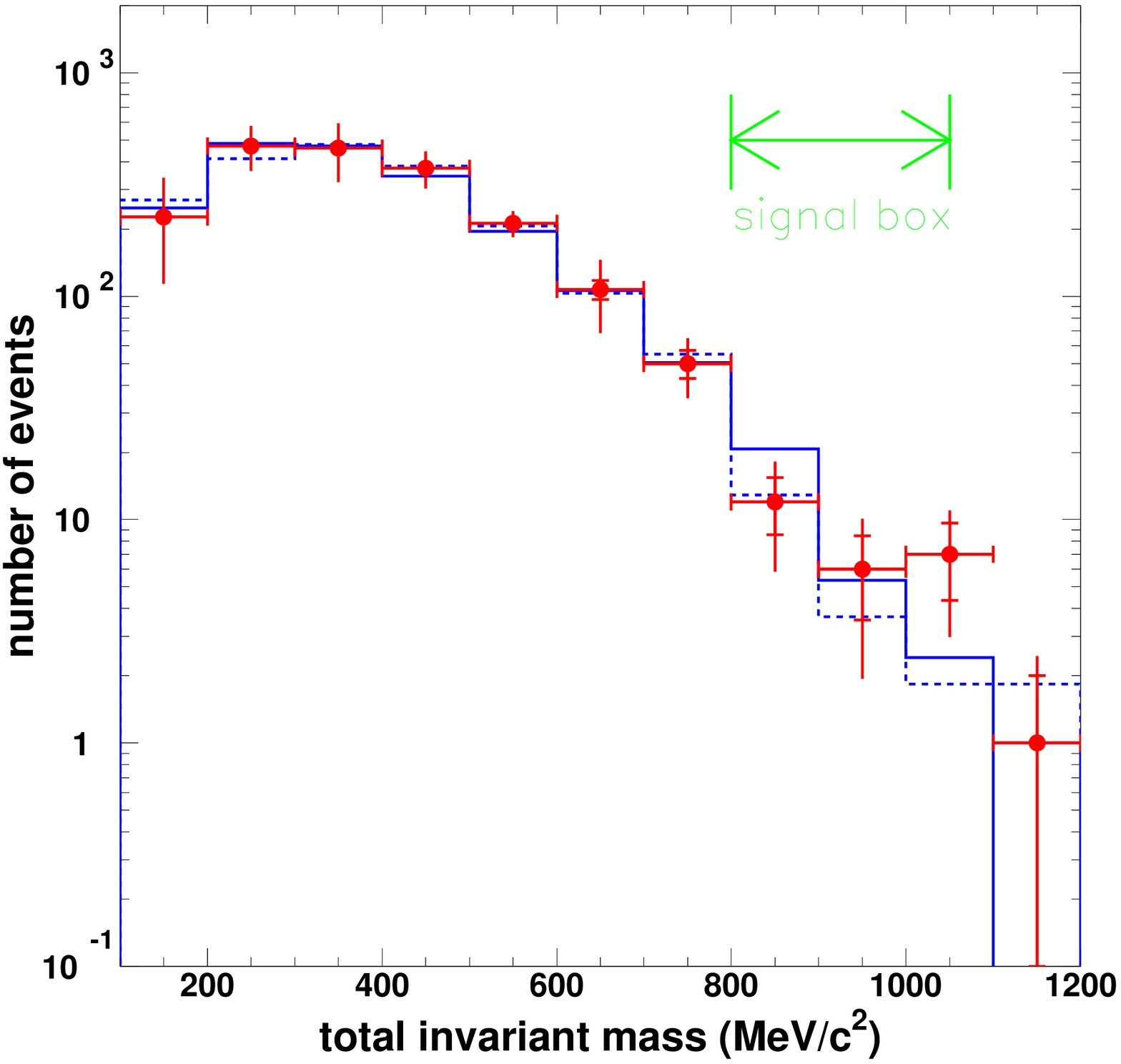}
          \caption
         {\protect \small Total invariant mass $M_{tot}$ for ``$\mu\pi^0$'' events
           with total momentum $P_{tot} \le 250~\hbox{MeV}/c$.
           Red crosses show the data with statistical and total measurement errors.
           The NEUT~(NUANCE) predictions are shown by a solid~(dashed) blue histogram.
           Both MC samples are normalized to the data by number of entries.
           The green arrow shows the range accepted by the proton decay signal box.}
          \label{kt_mtot}
          \end{figure}

\begin{figure}
          \includegraphics[width=8cm,clip]{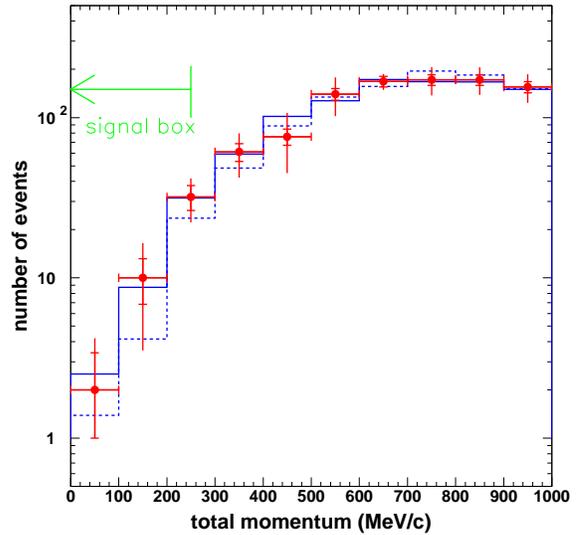}
          \caption
         {\protect \small Total momentum $P_{tot}$  for ``$\mu\pi^0$'' events
           with total invariant mass $800 \le M_{tot} \le 1050~\hbox{MeV}/c^2$.
           Red crosses show the data with statistical and total measurement errors.
           The NEUT~(NUANCE) predictions are shown by a solid~(dashed) blue histogram.
           Both MC samples are normalized to the data by number of entries.
           The green arrow shows the range accepted by the proton decay signal box.}
          \label{kt_ptot}
          \end{figure}

\begin{figure}
          \includegraphics[width=8cm,clip]{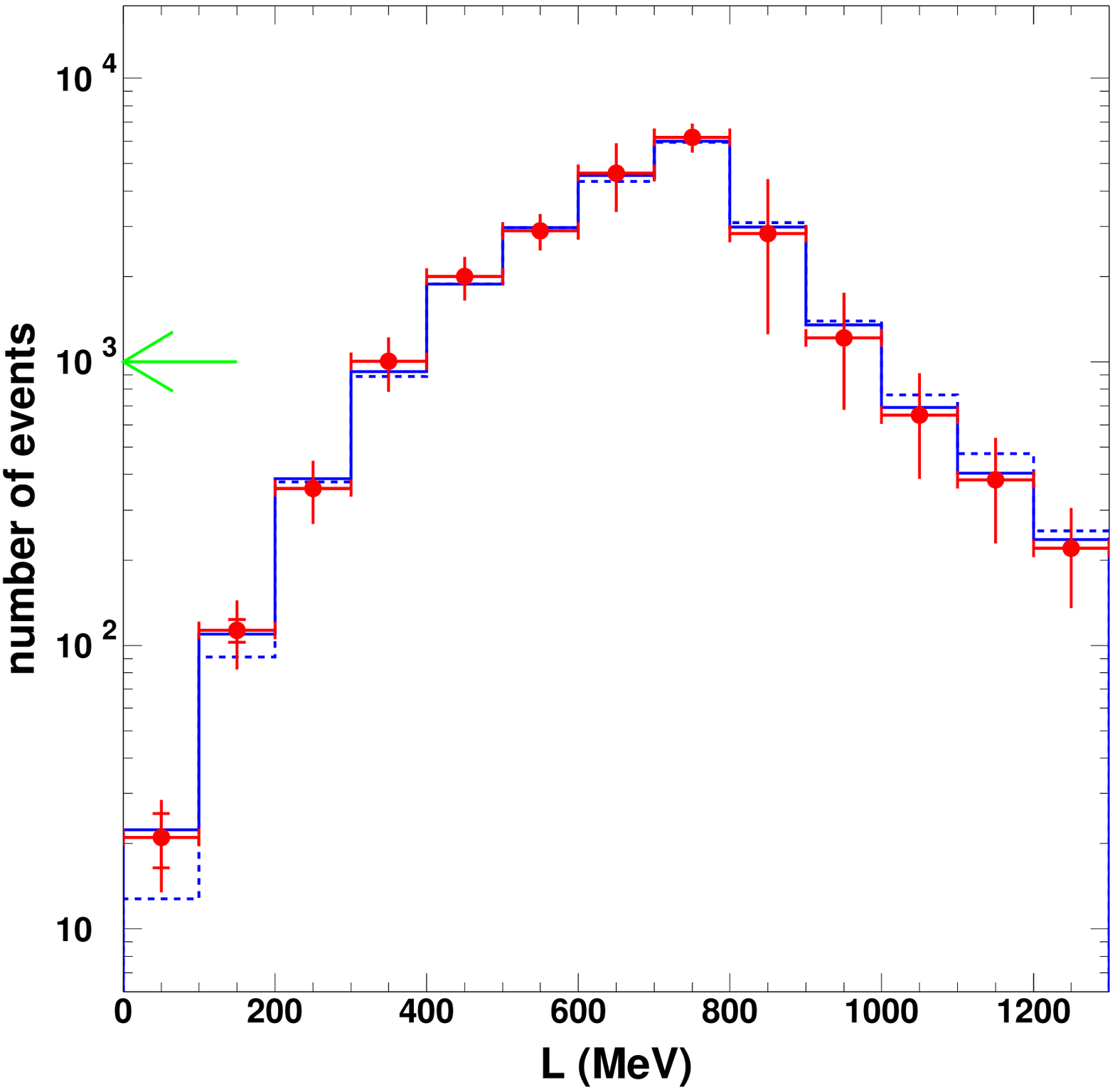}
          \caption
         {\protect \small $L$ (as defined in the text) for ``$\mu\pi^0$'' events.
           Red crosses show the data with statistical and total measurement errors.
           The NEUT~(NUANCE) predictions are shown by a solid~(dashed) blue histogram.
           Both MC samples are normalized to the data by number of entries.
           The green arrow indicates the range accepted by the proton decay signal box.}
          \label{kt_l}
          \end{figure}

To simplify comparison between data and MC samples, one dimensional
projections of $M_{tot}$ for $P_{tot} \le 250~\hbox{MeV}/c$ and
$P_{tot}$ for $800 \le M_{tot} \le 1050~\hbox{MeV}/c^2$ are shown in
Fig.~\ref{kt_mtot} and Fig.~\ref{kt_ptot}, respectively.  Note that omitted
statistical errors on the MC samples are almost the same as those on
data. Also, a metric quantifying an event's distance from the
center of the signal box is introduced: $L \equiv \sqrt{X^2+Y^2}$,
where
\begin{eqnarray*}
X & \equiv & M_{tot}c^2 - 938~\hbox{MeV} \\
Y & \equiv & \left \{
\begin{array}[c]{ll}
P_{tot}c - 200~\hbox{MeV} & (P_{tot} > 200~\hbox{MeV}/c) \\
0 & (P_{tot} \le 200~\hbox{MeV}/c)
\end{array}
\right.
. 
\end{eqnarray*}
The $L$ distributions for data and MC samples are plotted in Fig.~\ref{kt_l}.
The data and both simulations agree well, within measurement errors, 
for all values of $M_{tot}$~(Fig.~\ref{kt_mtot}),
$P_{tot}$~(Fig.~\ref{kt_ptot}), and $L$~(Fig.~\ref{kt_l}), 
including the tails of distributions 
where the proton decay signal box is located.


In summary, these comparisons show good agreement between data and MC 
samples; hence, the modeling of neutrino and final-state nuclear interactions relevant to
$p \rightarrow e^+ \pi^0$ backgrounds appears to be well supported by the KT data.

\subsection{Determination of the background rate to 
the $p\rightarrow e^+\pi^0$ search}

\subsubsection{General method}

Predicting the background rate for the $p \rightarrow e^+\pi^0$ search
requires extrapolation from the KT $\nu_\mu$ beam data to all flavors
of the atmospheric neutrino flux, and re-weighting the K2K neutrino
spectrum to match the atmospheric neutrino flux and spectrum.
The CC background event rate from $\nu_e$ is obtained using lepton
universality, the assumption that the cross sections and final state
kinematics for $\nu_e$ and $\nu_\mu$ are nearly identical.
The NC background rate is directly measured.  The anti-neutrino
background rate for all flavors is estimated by rescaling the
$\nu_\mu$ data to the predicted total event rates for each
anti-neutrino.  
Anti-neutrino final-states tend to have smaller $Q^2$
due to the different fraction of neutrino energy transferred
to the hadronic system in inelastic reactions
and therefore tend to have a larger momentum
imbalance between the lepton and hadronic systems.
Simulated data samples confirm that anti-neutrinos are less likely
than neutrinos to fall within the signal box, so the rescaling
procedure is conservative.

The expected event rate of the $p\rightarrow e^+\pi^0$ background for general 
water Cherenkov detectors, $N$, can be expressed as:
\begin{equation}
N = n^{CC} \cdot R_{\phi}^{CC} \cdot R_{\epsilon}^{CC}
+ n^{NC} \cdot R_{\phi}^{NC} \cdot R_{\epsilon}^{NC},
\label{eqn:measurement}
\end{equation}
where the 1st~(2nd) term corresponds to the background rate coming
from the CC~(NC) neutrino interactions, $n$ is the observed events in
the proton decay signal box at the KT, and $R_{\phi}$ is a ratio of
total neutrino interactions (flux $\times$ total cross section
$\times$ target volume $\times$ time) between the atmospheric
neutrinos at the proton decay detectors and the K2K muon neutrinos at
the KT.  
In using Eqn.~\ref{eqn:measurement}, $\nu_e$+$\bar{\nu}_e$ and
$\nu_{all}$+$\bar{\nu}_{all}$ (where $all$ stands for all the neutrino
flavors) are used for the atmospheric neutrino flux for the CC and NC
background measurements, respectively.  
$R_{\phi}$ is essentially an energy-dependent flux correction
factor, multiplied by an overall scale factor based on the respective
target masses and exposure.  Finally, $R_{\epsilon}$ is the ratio of
detection probabilities for the background events at the proton decay
detectors and the KT.  Any difference in detection efficiency between
the ``$p\rightarrow e\pi^0$'' sample in the proton decay detectors and
the ``$p\rightarrow \mu\pi^0$'' sample in the KT is reflected in
$R_{\epsilon}^{CC}$.

Since the numbers of "$p\rightarrow \mu\pi^0$" events from NEUT and 
NUANCE at KT are statistically consistent with each other, 
one (NEUT) of the models is chosen for the background rate determination
in this study.


As shown in Sec.~III-A-2, 
the remaining number of ``$\mu\pi^0$'' events
in the proton decay signal box, $n^{CC}$, is 24 for 7.4 $\times$
10$^{19}$~pot.  

The number of NC background events, 
the NC interactions with only two $\pi^0$s visible in the final state, 
is measured with the KT 
by applying the same event selection criteria used in the SK
$p\rightarrow e^+ \pi^0$ search.  
The events must be fully-contained,
and have two or three rings, all of which must be electron-like.  For
three-ring events, there must be a reconstructed $\pi^0$ mass between
85 and 215~MeV/$c$$^2$. 
The total mass of the event must satisfy the
conditions 800 $\le$ $M_{tot}$ $\le$ 1050 MeV/$c^2$, and the $P_{tot}$
$\le$ 250 MeV/$c$.  

In the SK data analysis, a decay-electron cut is also applied to the
data and MC samples. The reconstruction software used at the KT during
K2K running did not implement a decay-electron cut.  However, it was
found that the NC background rate was so small after the previous cuts that
there were no data candidates left to check for muon-decays. For the KT
MC ``$p\rightarrow e\pi^0$'' sample, 
a cut was applied using the MC truth information assuming the
same decay-electron finding efficiency as SK.  With the exceptions of
the decay-electron cut and method for FC event selection, which in SK
employs the outer detector, the selection criteria applied for these two
detectors are identical.

No candidates of the ``$p\rightarrow e\pi^0$'' events are found in the
KT data ($n^{NC}$ = 0 for 7.4 $\times$ 10$^{19}$~pot), while the
expected number of NC events is 0.30$\pm$0.13~(stat.) for
the same number of total neutrino interactions
in the 25~ton fiducial volume.
All the MC ``$p\rightarrow e\pi^0$'' events from the NC interactions 
are found to have only two $\pi^0$s visible in the final state.  These
events were NC single pion resonance events with an extra pion
produced by hadronic interactions in the water.

\subsubsection{Application to Super-Kamiokande-I}

\begin{figure}
  \includegraphics[width=8cm,clip]{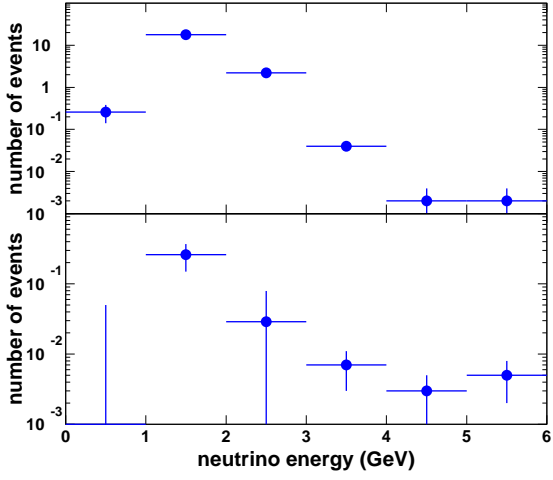}
  \caption
  {\protect \small Predicted number of CC ($m^{CC}(i)$ shown in top)
    and NC ($m^{NC}(i)$ shown in bottom) ``$p\rightarrow\mu(e)\pi^0$''
    events from NEUT in the KT for 7.4 $\times$ 10$^{19}$~pot, in bins of true
    parent neutrino energy $i$. The errors shown are MC statistical.}
  \label{masspro_mc_sb}
\end{figure}

In order to calculate the background to the proton decay search at the
SK-I experiment~\cite{sk_epi0_exp2,sk_epi0_exp3}  the interaction
rates are first re-weighted to the atmospheric flux in SK~\cite{honda,
  atmpd_full} and the SK fiducial volume of 22.5~kiloton and used to
calculate:
\begin{equation}
  R_{\phi}^{CC(NC)} = {\sum_i^{energy~bins} m^{CC(NC)}_{i} 
    \cdot r_{\phi \ i}^{CC(NC)} \over 
    \sum_j^{energy~bins} m^{CC(NC)}_{j}} ,
\label{eqn:fratio}
\end{equation}
where $m^{CC(NC)}$ is the expected number of
``$p\rightarrow\mu(e)\pi^0$'' events from CC(NC) interactions in the
KT for 7.4 $\times$ 10$^{19}$~pot and $r_{\phi}^{CC(NC)}$ is the
expected ratio of
$\nu_e$+$\bar{\nu}_e$~($\nu_{all}$+$\bar{\nu}_{all}$) interactions in
SK for 1~Mtyr to the $\nu_\mu$ interactions in the KT.
$r_{\phi}^{CC(NC)}$ and $m^{CC(NC)}$ 
are shown in Fig.~\ref{flux_comp}~(bottom) and Fig.~\ref{masspro_mc_sb}, respectively.

The background rate is estimated only for $E_{\nu} < 3 \, \hbox{GeV}$
in this study
since the expected number of ``$p\rightarrow\mu(e)\pi^0$'' events in
the KT is small~($<$10$^{-1}$) above 3~GeV as shown in
Fig.~\ref{masspro_mc_sb}.  The obtained results for $E_{\nu} < 3 \,
\hbox{GeV}$ are $R_{\phi}^{CC}$ = (15.9~Mtyr)$^{-1}$
and $R_{\phi}^{NC}$ = (4.5~Mtyr)$^{-1}$.
For example, $R_{\phi}^{CC}$ = (15.9~Mtyr)$^{-1}$ means 
that KT $\nu_{\mu}$ data 
corresponds to 15.9~Mtyr exposure of the atmospheric neutrino data of
the CC background events at SK.
The difference between $R_{\phi}^{CC}$ and $R_{\phi}^{NC}$ mostly comes 
from the difference between $r_{\phi}^{CC}$ and $r_{\phi}^{NC}$.
The fraction of background events above 3~GeV in a search for
$p\rightarrow e^+\pi^0$ at SK is reported to be about 24\,\% using
MC~\cite{future_wc1}.
%

The efficiency ratio for the background events, $R_{\epsilon}^{CC(NC)}$,
can be estimated by using proton decay signal MC
(note that the majority of the background events really have visible
Cherenkov rings and their kinematics are consistent with proton
decay signals as shown in Sec.~III-A-2):
$R_{\epsilon}^{CC(NC)}$ = $\varepsilon$ / $\epsilon^{CC(NC)}$,
where $\varepsilon = 0.40$ is the efficiency for
$p\rightarrow e^+\pi^0$ MC events at SK and
$\epsilon^{CC(NC)}=0.37(0.34)$ are the efficiencies for
$p\rightarrow$$\mu^+\pi^0$ and $p\rightarrow e^+\pi^0$ MC events
in the KT.  The largest source of inefficiency for both detectors
is due to final-state pion interactions in the oxygen nucleus.
The efficiencies for free protons are 0.80 and 0.77 for
$p\rightarrow e^+\pi^0$ MC events at SK and $p\rightarrow$$\mu^+\pi^0$ MC 
events in the KT, respectively.
The 15\,\%~(=0.34/0.40) difference between SK and KT
efficiencies for $p\rightarrow e^+\pi^0$ is mainly due to the different
FC selection criterion.

Finally, using the numbers above the atmospheric neutrino background
rate for the SK $p\rightarrow e^+\pi^0$ search, $N$, is
determined to be:
\begin{eqnarray*}
N~(Mtyr^{-1}) = \{1.63 \pm 0.33 (stat.) ^{+0.43}_{-0.51} (syst.)\} (CC) \nonumber \\
+ \{0.00 + 0.26 (stat.) + 0.13 (syst.)\} (NC) \nonumber \\
= 1.63 ^{+0.42}_{-0.33} (stat.) ^{+0.45}_{-0.51} (syst.). \nonumber
\end{eqnarray*}

Table~\ref{syserr} summarizes the systematic errors on each parameter
in Eqn.~(\ref{eqn:measurement}).
\begin{table}
  \begin{center}
    \begin{tabular}{lr}
      \hline \hline
      Error source & error on $n^{CC}$~(\%)\\
      \hline
      KT vertex reconstruction & 3\\
      KT absolute energy scale & 3\\
      KT detector asymmetry of energy scale & 3\\
      KT FC event selection & 1\\
      KT ring counting & 9\\
      KT ring direction & 3\\
      KT PID & 9\\
      KT NC fraction in ``$p\rightarrow\mu\pi^0$'' events & +0/-4\\ 
      \hline
      Sub total & +14/-15\\
      \hline \hline \\
      \hline \hline
      Error source  & error on $R_{\phi}^{CC(NC)}$~(\%)\\
      \hline
      K2K flux shape & 3(6)\\
      K2K beam $\nu_e$ contamination & $<$1(5)\\
      KT MC normalization & 4(4)\\
      KT MC statistics & 5(47)\\
      atmospheric $\nu$ flux~\cite{honda, atmpd_full} & 10(10)\\
      total $\nu$ cross section ratio & 15(15)\\
      lepton universality & $<$1(0)\\
      atmospheric anti-$\nu$ rate & +0/-16(6)\\
      \hline
      Sub total & +25/-29(54)\\
      \hline \hline \\
      \hline \hline
      Error source  & error on $R_{\epsilon}^{CC(NC)}$~(\%)\\
      \hline
      Efficiency difference between SK and the KT & 11(7)\\
      \hline \hline
    \end{tabular}
  \end{center}
  \caption{\protect \small
    Summary of systematic errors for measurement of the
    background rate to the $p\rightarrow e^+\pi^0$ search.}
  \label{syserr}
\end{table}
Uncertainties from the performance of the KT reconstruction are reflected in
the error on $n^{CC}$.
The error is estimated by evaluating the effect of reconstruction
uncertainties on the number of $p\rightarrow\mu^+\pi^0$ MC events in
the proton decay signal box.  To account for the NC contamination in
the ``$p\rightarrow\mu\pi^0$'' CC events~(Tbl.~\ref{neutmode}), 
an asymmetric error of~(+0/-4\,\%)
is assigned to $n^{CC}$. 
It was not necessary to estimate the errors on $n^{NC}$ as there was no
observation of ``$p\rightarrow e\pi^0$'' events in the KT.  
The expected event rate in this case is much smaller than that 
from the CC events, which dominates the background.
%
%

One (15\,\%) of the dominant errors on $R_{\phi}^{CC}$ comes from
uncertainty of total neutrino interactions with atmospheric neutrino
flux at SK.
The cross section systematic errors considered are
10\,\% on the axial vector masses for quasi-elastic and single-pion production,
10\,\% on the total cross sections for quasi-elastic and single-pion production,
5\,\% on the total cross section for deep-inelastic scattering,
with and without correction of the nuclear structure functions
for deep-inelastic scattering~(Bodek \& Yang correction~\cite{bodek}),
and 20\,\% on the ratio of NC/CC cross sections.
The other dominant error (+0/-16\,\%) on $R_{\phi}^{CC}$ comes from
differences in kinematics of the final-state particles between
neutrinos and anti-neutrinos. 
The error on $R^{CC}_{\phi}$~($R^{NC}_{\phi}$) is estimated
by comparing the number of simulated CC single-$\pi^0$
(NC two-$\pi^0$) events near the signal box.
The error on $R^{CC}_{\phi}$ from the lepton universality assumption
is estimated by comparing the number of CC single-$\pi^0$ events near
the signal box using simulated $\nu_{\mu}$ and $\nu_e$ interactions
and is found to be negligibly small.
%
%
%
%
%
The largest error~ (47\,\%) on $R_{\phi}^{NC}$ 
arises from the statistics of simulated ``$p\rightarrow e\pi^0$'' events
($m^{NC}$ in Eqn.~(\ref{eqn:fratio})) in the KT.
The KT MC data are normalized by the number of total neutrino interactions 
observed in the 25~ton fiducial volume.
The estimate of the error on the fiducial volume is 4\,\%~\cite{k2k_full}.

The systematic error on $R_{\epsilon}^{CC(NC)}$ is estimated by taking
into account the differences of PID performance between SK and the KT
and the mis-identification probability of charged pions or protons as
either muons or gammas in the KT MC sample.


\subsubsection{Discussions}

A total of 2.14 (=1.63/0.76, where 0.76 is fraction of the background events 
below 3~GeV at SK\cite{future_wc1}) atmospheric neutrino background events
are expected for 1~Mtyr for SK-type water Cherenkov detectors.
If the proton lifetime $\tau/B_{p\rightarrow e^+\pi^0}$ were 6$\times$10$^{34}$
years, the proton decay signal rate would be almost equivalent to the rate of 
the background events.
Therefore, it will be important to reduce the number of the background events 
in order to make a clean discovery of proton decay in future experiments.
If necessary, the remaining background can be further reduced to improve 
the signal/background ratio.
For example, Fig.~\ref{kt_ptot} suggests that the background rate
could be reduced by an order of magnitude
by applying a tighter momentum cut~($P_{tot}<$100~MeV/$c$),
while the signal efficiency would be reduced only by a factor of 2.3
\cite{future_wc1}.

With the present SK exposure~(0.092~Mtyr),
the expected background rate for $E_{\nu} < 3 \,\hbox{GeV}$ is 
0.15 $^{+0.04}_{-0.03}$~(stat) $^{+0.04}_{-0.05}$~(syst) events.
This result is consistent with the observed lack of candidates
in the SK experiment~\cite{sk_epi0_exp2, sk_epi0_exp3}.

According to Tbl.~\ref{syserr}, the background rate determination can
be improved by reducing uncertainties of the atmospheric neutrino flux,
and uncertainties of total and anti-neutrino interaction rates with the 
atmospheric neutrino flux in future. 

Note that the background rate for the $p\rightarrow \mu^+\pi^0$ mode
can be determined in the same way just by replacing the atmospheric
$\nu_e$+$\bar{\nu}_e$ flux with $\nu_\mu$+$\bar{\nu}_\mu$ flux
for the CC background and the detection probability of the
``$p\rightarrow e\pi^0$'' sample 
with that of the ``$p\rightarrow \mu\pi^0$'' sample
in the proton decay detectors.

\section{Conclusion}

The atmospheric neutrino background to searches for $p \rightarrow e^+
\pi^0$ has been experimentally studied using an accelerator neutrino
beam and a water Cherenkov detector for the first time.  
The K2K KT detector, with the same target material and detection technique 
as SK, accumulated data equivalent to atmospheric neutrino exposures of 15.9
and 4.5~Mtyr for the CC and NC background events, respectively.

Two neutrino interaction simulation programs,
NEUT and NUANCE, were evaluated and found
to reproduce accelerator neutrino beam interactions in water within
the measurement errors.  Both should therefore reliably model the
atmospheric neutrino background to $p\rightarrow e^+ \pi^0$ searches.

This is the first determination of the background rate using data
for the SK-type water Cherenkov detectors.
We measured the rate of neutrino and final-state nuclear interactions of the
background events.
To measure the background rate from the KT data themselves, almost
identical selection criteria as for $p \rightarrow e^+ \pi^0$ searches
were applied.  A total of 24 ``proton decay''-like events were
identified.  A re-weighting procedure was applied using the
atmospheric neutrino flux and detection efficiency~(40\,\%) for the SK proton
decay search, and the resulting background rate for $E_{\nu} < 3 \,
\hbox{GeV}$~(where about 76\,\% of the background events
is expected in SK according to the MC study) 
is $1.63 ^{+0.42}_{-0.33}\,\hbox{(stat)}
^{+0.45}_{-0.51}\,\hbox{(syst)}\, \hbox{Mtyr}^{-1}$.  

%
This experimentally determined background rate is consistent with no
candidates being observed over the 0.1~Mtyr exposure of the SK-I experiment.

This result shows that about two background events per year would be
expected in possible future one-megaton-scale detectors.  Assuming a
finite proton lifetime by an order beyond the present limit, the
rate of signal would be similar to the expected background rate, both
in these proposed detectors as well as in a still-running
Super--Kamiokande.  Therefore, further reduction of
the background events will be crucial in the future.
%

\section{Acknowledgments}

We thank the KEK and ICRR directorates for their strong support and
encouragement.  K2K was made possible by the inventiveness and the
diligent efforts of the KEK-PS machine group and beam channel group.
We gratefully acknowledge the cooperation of the Kamioka Mining and
Smelting Company.  This work has been supported by the Ministry of
Education, Culture, Sports, Science and Technology of the Government
of Japan, 
the Japan Society for Promotion of Science, the U.S. Department of
Energy, the Korea Research Foundation, the Korea Science and
Engineering Foundation, NSERC Canada and Canada Foundation for
Innovation, the Istituto Nazionale di Fisica Nucleare (Italy), 
the Ministerio de Educaci\'on y Ciencia and Generalitat Valenciana (Spain),
the Commissariat \`{a} l'Energie Atomique (France), and Polish KBN grants:
1P03B08227 and 1P03B03826.



\end{document}